\documentclass[preprint,prc,nofootinbib,superscriptaddress,a4]{revtex4}
\usepackage{color}
\usepackage{amssymb}
\usepackage{ulem}
\usepackage{amsmath}
\usepackage{graphicx}

\usepackage{ulem}
\usepackage{xcolor}
\def\be{\begin{equation}} \def\ee{\end{equation}} \def\bea{\begin{eqnarray}}
\def\eea{\end{eqnarray}} 

\def\te{\color{teal}}

\begin{document}

\title{
Joint cluster-EFT analysis of $^{16}$N $\beta$-delayed $\alpha$ spectra 
and $\alpha$-$^{12}$C scattering
}

\author{Jubin Park}
\affiliation{Department of Physics and Origin of Matter and Evolution of Galaxies Institute,
Soongsil University, Seoul 06978, Korea}

\author{Myeong-Hwan Mun}
\email{aa3101@gmail.com}
\affiliation{Department of Physics, Kyungpook National University, Daegu 41566, Korea}

\author{Shung-Ichi Ando}
\affiliation{Department of Display and Semiconductor Engineering and Research Center for Nano-Bio Science, Sunmoon University, Asan 41439, Korea}

\date{\today}

\begin{abstract}
The $\beta$-delayed $\alpha$ decay (BDAD) of $^{16}$N probes the
$p$-wave $\alpha$-$^{12}$C continuum relevant to the low-energy
$E1$ 
transition of $^{12}$C$(\alpha,\gamma)^{16}$O.
We analyze the spectra of Azuma \textit{et al.} and Tang
\textit{et al.} within cluster effective field theory, fitting each
independently together with the same elastic-scattering data.
The baseline eight-parameter fits yield $\chi^2_{\rm A}/N_{\rm A} = 1.732$ for the Azuma spectrum and $\chi^2_{\rm T}/N_{\rm T} = 2.028$ for the Tang spectrum, substantially improving upon the corresponding fixed-propagator values of 4.06 and 3.56, respectively.
The raw simultaneous fit yields
$\chi^2_{\rm A}/N_{\rm A} = 4.319$, $\chi^2_{\rm T}/N_{\rm T} = 10.901$, and $\chi^2_{\rm el}/N_{\rm el}= 6.366$ for the Azuma, Tang, and elastic-scattering data, respectively. With sector-balanced weighting, the corresponding values are 3.308, 8.993, and 6.466, respectively.
Thus, reducing the relative elastic weight improves the BDAD sectors but
does not recover the quality of the independent fits.
The common strong parameters and normalization ratio remain stable,
whereas the fitted weak-current coefficients depend on the objective.
The two spectra are separately compatible with a common strong continuum,
but the minimal common weak-current amplitude does not reproduce them
simultaneously; its fitted coefficients also depend on the objective.
These results define the experimental and model sensitivities 
that must be propagated in a future unified analysis including radiative capture.
\end{abstract}

\maketitle

\section{Introduction}
\label{sec:introduction}

The $^{12}$C$(\alpha,\gamma)^{16}$O reaction is one of the principal
reactions governing stellar helium burning~\cite{f-rmp84,rr-88}.
Together with the triple-$\alpha$ process, it determines the
carbon-to-oxygen ratio after core helium exhaustion and thereby affects
subsequent stellar evolution and nucleosynthesis.
The reaction rate is required near the Gamow energy,
$E_{\rm cm}\simeq0.3$~MeV, where Coulomb suppression makes a direct
measurement impractical.
It must therefore be inferred by extrapolating higher-energy capture data
with complementary constraints from elastic scattering, transfer reactions,
and nuclear decay~\cite{deboer,metal-prc24,detal-epja25}.
Controlling this extrapolation and its uncertainty remains a central problem
in nuclear astrophysics.

The $\beta$-delayed $\alpha$ decay (BDAD) of $^{16}$N provides an important
indirect probe of the low-energy $E1$ capture amplitude~\cite{bb-npa06}.
It populates the $p$-wave $\alpha$-$^{12}$C continuum in $^{16}$O,
including the region influenced by the subthreshold $1^-_1$ state only
45~keV below the $\alpha$ threshold.
The same strong $p$-wave pole and continuum dynamics enter
$^{12}$C$(\alpha,\gamma)^{16}$O, although the weak and
electromagnetic production operators are different.
The probing $\beta$-decay phase space toward low $\alpha$ energy
enhances the sensitivity of the decay spectrum to the subthreshold
region.

The low-energy structure of the BDAD of $^{16}$N was first observed with sufficient sensitivity
in the Yale measurement~\cite{Zhao1993}, and the high-statistics
coincidence spectrum of Azuma \textit{et al.} subsequently became a
standard input in analyses of the $E1$ strength~\cite{Azuma1994}.
Later measurements and reanalyses revealed differences in the main-peak
width and interference minimum, which affect the separation of the
$p$- and $f$-wave contributions~\cite{France2007}.
Tang \textit{et al.} obtained an independent coincidence spectrum extending
to an $\alpha$ energy of 450~keV using twin ionization chambers
~\cite{tang}.
Kirsebom \textit{et al.} subsequently reduced the absolute-normalization
uncertainty through improved $^{16}$N branching ratios~\cite{Kirsebom2018}.
Nevertheless, the extent to which the Azuma and Tang spectra imply
the same underlying decay amplitude has remained unresolved.

Most quantitative analyses of this system 
use the $R$-matrix formalism~\cite{lt-rmp58,v-rmp62,db-rpp10},
which provides a successful description of elastic scattering, radiative
capture, and $\beta$-delayed decay~\cite{deboer}.
Cluster effective field theory (EFT) offers a complementary organization
that can test the parametrization dependence of the inferred continuum
dynamics~\cite{w-pa79,hkvk-rmp20}.
In cluster EFT, unresolved short-distance physics is absorbed into local
operators, while the strong $\alpha$-$^{12}$C interaction is encoded in
dressed partial-wave $^{16}$O propagators common to scattering, capture,
and decay~\cite{sa-epja21}.
The BDAD line shape 
is generated by pole and non-pole weak amplitudes built
on these propagators, rather than by assigning an independent set of level
parameters to each reaction.
A joint analysis can therefore test directly whether observables generated
by different external probes are consistent with the same low-energy strong
amplitude.

A quantitative basis for the present analysis was established in our recent
cluster-EFT study of elastic $\alpha$-$^{12}$C scattering~\cite{mpha-prc26}, in which $11\,392$ differential-cross-section
measurements~\cite{tetal-prc09} were used to determine the common
strong-sector parameters.
Earlier calculations of the $^{16}$N spectra kept this propagator fixed and obtained 
$\chi^2_{\rm A}/N_{\rm A} = 4.06$ for the Azuma spectrum 
and $\chi^2_{\rm T}/N_{\rm T} = 3.56$ for the Tang spectrum~\cite{sa-epja21}, 
indicating that the fixed-propagator
description was qualitatively successful but quantitatively insufficient.

The numerical strategy follows our elastic-scattering analysis~\cite{mpha-prc26}, 
in which differential evolution (DE) was used to locate
the global 37-parameter elastic solution and ensemble Markov-chain Monte
Carlo (MCMC) sampling was used to quantify parameter correlations and
uncertainties.
In the present work, independent DE searches first identify a reproducible
best-fit solution for each objective, and the selected solution then
initializes ensemble MCMC sampling of the corresponding parameter
distribution.
The same DE--MCMC sequence is applied to the independent and simultaneous
analyses; further details are given in Sec.~\ref{subsec:protocol}.

Here, each BDAD spectrum is first fitted independently together with the
same elastic data using a minimal eight-parameter amplitude.
The resulting fits give 
$\chi^2_{\rm A}/N_{\rm A} = 1.732$ for the Azuma spectrum
and $\chi^2_{\rm T}/N_{\rm T} = 2.028$ for the Tang spectrum,
corresponding to 
reductions in $\chi^2/N$
of approximately 57\% and 43\% over the
fixed-propagator results.
The shared $p$-wave effective-range parameters remain close to the elastic
reference solution, showing that both spectra are separately compatible
with essentially the same strong $\alpha$-$^{12}$C continuum.
No additional effective $1^-$ resonance or 
phenomenological high-energy background pole
is introduced in this baseline description.

We then test the mutual compatibility of the two measurements by imposing
common strong and weak-current coefficients while retaining independent
overall normalizations. 
The raw simultaneous fit yields $\chi^2_{\rm A}/N_{\rm A} = 4.319$, $\chi^2_{\rm T}/N_{\rm T} = 10.901$, and $\chi^2_{\rm el}/N_{\rm el} = 6.366$ for the Azuma, Tang, and elastic-scattering data, respectively, whereas the sector-balanced simultaneous fit yields $\chi^2_{\rm A}/N_{\rm A} = 3.308$, $\chi^2_{\rm T}/N_{\rm T} = 8.993$, and $\chi^2_{\rm el}/N_{\rm el} = 6.466$.
Reducing the relative elastic weight improves both BDAD sectors, but neither approaches the quality of its independent fit. The common-current  
tension is therefore not an artifact of the much larger number of elastic data points.
The independently fitted normalization coefficients $C_{\mathrm{OAC}}^{A}$ and $C_{\mathrm{OAC}}^{T}$, associated with the Azuma and Tang spectra, respectively, absorb the overall scale difference between the two data sets. Their ratio, $C_{\mathrm{OAC}}^{A}/C_{\mathrm{OAC}}^{T}$, is 3.7489 for the raw fit and 3.7425 for the sector-balanced fit.
The normalization ratio and the shared strong-sector parameters are stable
under the change of objective, whereas the common weak-current coefficients
are not.  
Thus, within the present minimal common-current  
model and without
experiment-specific calibration, resolution, or response parameters,
the two published spectra are separately compatible with a common elastic
strong sector, but the minimal common weak-current amplitude does not
reproduce them simultaneously; its fitted coefficients also depend on
the objective.
This compatibility test identifies the decay-specific and experimental
line-shape ambiguities that must be propagated in a future unified analysis of
elastic scattering, BDAD, and radiative capture.

We do not extract a new value of $S_{E1}$(300$\,\mathrm{keV}$) here, because the
radiative-capture amplitude contains electromagnetic short-distance
couplings that are not fixed by elastic scattering or BDAD alone.
Instead, the present analysis establishes the common strong-interaction
basis and identifies the experimental and weak-sector ambiguities that
must be propagated to the capture channel.
A future unified cluster-EFT analysis~\cite{UnifiedCaptureWIP} of elastic scattering, both
BDAD spectra, and the $E1$ component of
$^{12}$C$(\alpha,\gamma)^{16}$O will test the consistency of the
strong, weak, and electromagnetic sectors and quantify their
correlated impact on $S_{E1}(E)$ in the helium-burning Gamow window.

The paper is organized as follows.
Section~II presents the cluster-EFT formalism for elastic $\alpha$-$^{12}$C scattering and the BDAD of $^{16}$N.
Section~\ref{sec:datafit} describes the experimental data sets, objective
functions, and DE--MCMC fitting procedure.
Section~\ref{sec:results} presents the independent eight-parameter analyses
and the simultaneous Azuma--Tang--elastic compatibility test.
The physical interpretation of these results and their implications for radiative capture are discussed in Section~\ref{sec:discussion}.
Finally, Section~\ref{sec:summary} summarizes the main conclusions and outlook.
The $^{16}$O state content and the posterior and numerical diagnostics
are summarized in Appendices A and B, respectively.

\section{Cluster EFT framework}
\label{sec:formalism}

In this section, we summarize the cluster-EFT framework used in the present work. 
Our focus here is restricted to the common $\alpha$-$^{12}$C sector~\cite{sa-prc23,mpha-prc26} 
and to the weak
operators relevant to the BDAD of $^{16}$N~\cite{sa-epja21}. 
The fitting strategy and the hierarchy of working parametrizations are deferred to the next section.

The central idea of the present approach is that the low-energy dynamics of the
$\alpha$+$^{12}$C system are encoded in dressed $^{16}$O propagators which are shared by
elastic scattering and by the BDAD of $^{16}$N.
In cluster EFT, the ground states of $\alpha$ and $^{12}$C are treated as effective pointlike
degrees of freedom, while higher excited configurations are integrated out and absorbed into
short-range operators. 
Because the experimental data used to determine the $\alpha$-$^{12}$C interaction lie at
energies above the strict low-energy expansion scale of the BDAD of $^{16}$N,
the combined study of the $\alpha$-$^{12}$C sector and the $^{16}$N weak decay sector 
may restrict the parameter space at low energy, which is essential when one extrapolates the $S$ factor
of the $^{12}$C($\alpha$,$\gamma$)$^{16}$O to the Gamow peak energy, $E_G\simeq 300$~keV. 

\subsection{Common $\alpha$-$^{12}$C sector}
\label{subsec:strong}

Restricting the EFT to the sectors relevant for the present work, we write the effective
Lagrangian schematically as~\cite{sa-epja21}
\begin{equation}
\mathcal{L}
=
\mathcal{L}_{\mathrm{ES}}
+
\mathcal{L}_{\beta N},
\label{eq:L_total}
\end{equation}
where $\mathcal{L}_{\mathrm{ES}}$ describes 
the elastic $\alpha$-$^{12}$C scattering at low energies and
$\mathcal{L}_{\beta N}$ contains the weak operators for the $^{16}$N decay channel.

The strong-interaction part is organized in terms of 
the $\alpha$ field $\phi_\alpha$ and $^{12}$C field $\phi_C$, 
and composite fields $d_{(li)}$ 
for the $i$th excited or resonant state of $^{16}$O
carrying definite orbital angular momentum $l$~\footnote{
The composite $^{16}$O states represented in the elastic reference
amplitude and their roles in the present fit are summarized in
Appendix~\ref{app:state_content}.}~\cite{sa-prc23}:
\begin{align}
\mathcal{L}_{\rm ES}
={}&
\phi_\alpha^\dagger
\left(
iD_0+\frac{\bm D^2}{2m_\alpha}
\right)\phi_\alpha
+
\phi_C^\dagger
\left(
iD_0+\frac{\bm D^2}{2m_C}
\right)\phi_C
\nonumber\\
&+
\sum_l\sum_i\sum_{k=0}^{n_l}
C_{(li)k}\,
d_{(li)}^\dagger
\left[
iD_0+\frac{\bm D^2}{2(m_\alpha+m_C)}
\right]^k
d_{(li)}
\nonumber\\
&-
\sum_l\sum_i y_{(li)}
\left[
(\phi_\alpha O_l\phi_C)^\dagger d_{(li)}
+
d_{(li)}^\dagger(\phi_\alpha O_l\phi_C)
\right]
+\cdots ,
\label{eq:Lstrong_new}
\end{align}

where $n_l = 3$ for $l = 0, 1, 2$ and $n_l = 4$ for $l = 3$.
The covariant derivative $D_\mu$ includes the Coulomb interaction,
$O_l$ projects the $\alpha$-$^{12}$C pair onto the $l$th partial wave,
and the ellipsis denotes higher-order short-range operators.
$m_\alpha$ and $m_C$ are the masses of $\alpha$ and $^{12}$C, respectively.
The form of the operators, e.g., $i D_0 + \bm{D}^2/2(m_\alpha + m_C)$, 
is invariant under the Galilean transformation, and in the center-of-mass frame, 
the total momentum of $d_{(li)}$ fields vanishes. 
Its powers generate polynomials through
$E^3$ ($p^6$) for $l=0,1,2$ and through
$E^4$ ($p^8$) for $l=3$.

The coefficients $C_{(li)k}$ are matched to the effective range parameters, or the energy, width, and shape
parameters of resonant states. 
The coefficients $y_{(li)}$ are redundant in the elastic scattering amplitudes.  
For the present BDAD analysis, the relevant continuum channels are the $p$-wave ($l=1$) and
the $f$-wave ($l=3$), corresponding to the dominant $1^-$ and $3^-$ contributions in the
$\alpha$+$^{12}$C final state.

\begin{figure}
\centering
\includegraphics[width=0.8\textwidth] 
{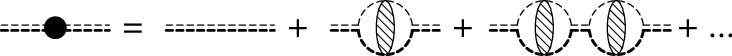}
\caption{
Feynman diagrams for dressed $^{16}$O propagators.
A thin-thick double dashed line with and without a filled circle denotes a dressed and bare
composite field, respectively. A thin (thick) dashed line is an $\alpha$ ($^{12}$C) field, and 
a shaded blob represents the Coulomb Green's function. 
}
\label{fig:16propagators}
\end{figure}

\begin{figure}
\centering
\includegraphics[width=0.2\textwidth]
{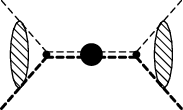}
\caption{
Feynman diagram for scattering amplitudes.
Shaded blobs represent the initial and final Coulomb wavefunctions.
The remaining notation is the same as in Fig.~\ref{fig:16propagators}.
}
\label{fig:scatteing_amplituds}
\end{figure}

The Coulomb-modified scattering amplitudes for the excited states of $^{16}$O are calculated
from the Feynman diagrams displayed in Figs.~\ref{fig:16propagators} and \ref{fig:scatteing_amplituds}.
The inverse dressed propagators are calculated from the diagrams in Fig. \ref{fig:16propagators} as
\begin{equation}
D_l(p) = K_l(p) - 2\kappa H_l(p) ,
\label{eq:Dl_new}
\end{equation}
with
\begin{equation}
H_l(p) = W_l(p)\,H(\eta),
\qquad
H(\eta) = \psi(i\eta) + \frac{1}{2i\eta} - \ln(i\eta),
\label{eq:Hdef_new}
\end{equation}
where $p=\sqrt{2\mu E}$ is the magnitude of center-of-mass momentum,
$\mu$ is the reduced mass of the $\alpha$-$^{12}$C system,
$\eta=\kappa/p$ is the Sommerfeld parameter; 
$\kappa$ is the inverse of the Bohr radius, $\kappa = \alpha_E\,\mu\, Z_\alpha Z_C$, where 
$\alpha_E$ is the fine structure constant, and $Z_\alpha$ and $Z_C$ are the numbers of protons
in the nuclei. 
$W_l(p)$ is the usual Coulomb polynomial factor: $W_l(p) = (\kappa^2/l^2 + p^2)W_{l-1}(p)$ 
with $W_0(p) = 1$. 
$\psi(z)$ is the digamma function. 

The short-range interaction is encoded in the effective-range function
\begin{equation}
K_l(p)
=
-\frac{1}{a_l}
+\frac{1}{2}r_l\, p^2
-\frac{1}{4}P_l\, p^4
+Q_l\, p^6
-R_l\, p^8
+\cdots .
\label{eq:Kl_new}
\end{equation}

As emphasized in the recent elastic-scattering EFT analyses~\cite{sa-prc18}, 
the strong Coulomb suppression
requires modified counting rules: terms up to $p^6$ are retained for $l=0,1,2$, whereas the
$l=3$ channel requires terms up to $p^8$.
The subthreshold bound states are incorporated through the pole condition
\begin{equation}
D_l(i\gamma_l)=0,
\qquad
\gamma_l=\sqrt{2\mu B_l},
\label{eq:pole_new}
\end{equation}
where $B_l$ is the corresponding binding energy.
We fix the scattering-length parameters $a_l$
by imposing the corresponding pole conditions and obtain
\bea
K_l(p) &=& \frac12 r_l(\gamma_l^2 + p^2) 
+ \frac14P_l (\gamma_l^4 - p^4)
+ Q_l(\gamma_l^6 + p^6) 
+ R_l(\gamma_l^8 - p^8) 
+ 2\kappa H_l(i\gamma_l) \,. 
\eea

The corresponding Coulomb-modified amplitudes are calculated from the diagrams in Fig. \ref{fig:scatteing_amplituds}, which may be written as
\begin{equation}
\tilde{A}_l(E)
=
\frac{C_\eta^2\, W_l(p)}
     {K_l(p)-2\kappa H_l(p)} ,
\label{eq:Aelastic_new}
\end{equation}
with
\begin{equation}
C_\eta^2 = \frac{2\pi\eta}{e^{2\pi\eta}-1}.
\label{eq:Ceta_new}
\end{equation}

Once the pole position is fixed, the asymptotic normalization coefficient (ANC) follows from the
derivative of the inverse propagator~\cite{irs-prc84},
\begin{equation}
|C_b|_l
=
\frac{\gamma_l^l}{l!}\,
\Gamma\!\left(l+1+\frac{\kappa}{\gamma_l}\right)
\left[
\left|
\frac{dD_l(p)}{dp^2}
\right|_{p^2=-\gamma_l^2}
\right]^{-1/2}.
\label{eq:ANC_new}
\end{equation}
The ANC magnitude $\left|C_b\right|_l$ in Eq.~\eqref{eq:ANC_new} 
should not be confused with the
weak-current pole coefficients $C_b^{(1)}$ and $C_b^{(3)}$ introduced
below.
In the present work, the dressed $p$- and $f$-wave propagators obtained from this common
$\alpha$-$^{12}$C sector are used directly in the BDAD amplitude.

A key conceptual point of the cluster-EFT description is that the subthreshold $1^-_1$ state
and the broad $1^-_2$ continuum strength are not treated as independent interfering levels in the
same way as in conventional $R$-matrix analyses.
Instead, the low-energy $p$-wave continuum is organized through a single dressed propagator,
and the observed line shape in the decay spectrum emerges from the structure of the full EFT amplitude
built on top of it.

\subsection{Weak decay amplitude for $^{16}$N($\beta\alpha$)$^{12}$C}
\label{subsec:weak}
The \(^{16}\mathrm{N}\) ground state has \(J^\pi=2^-\). The \(\beta\) transitions populating the \(1^-\) and \(3^-\) channels relevant to \(\beta\)-delayed \(\alpha\) emission are allowed Gamow--Teller transitions.
For the $\beta$-delayed $\alpha$ channel, the dominant $\alpha$-$^{12}$C final-state
configurations are the $p$- and $f$-waves.
We therefore construct the weak sector by coupling the $^{16}$N source field to the
$\alpha$-$^{12}$C continuum through an external axial current.

The weak interaction Lagrangian, including the leading operators and
the first momentum-dependent $p$-wave corrections, is written as~\cite{sa-epja21}
\begin{align}
\mathcal{L}_{\beta N}
={}&
C_a^{(1)} y_1
\bigl[
a_i (\phi_\alpha O_{1,j}\phi_C)
\bigr]^\dagger
\phi_{N,ij}
+
C_b^{(1)}
(a_i d_{1,j})^\dagger
\phi_{N,ij}
\nonumber\\[0.3em]
&+
C_a^{(3)} y_3
\bigl[
(\phi_\alpha O_{3,ijk}\phi_C)a_k
\bigr]^\dagger
\phi_{N,ij}
+
C_b^{(3)}
(d_{3,ijk}a_k)^\dagger
\phi_{N,ij}
\nonumber\\[0.3em]
&+
D_a^{(1)} y_1
\bigl[
a_i (\phi_\alpha O_1^2 O_{1,j}\phi_C)
\bigr]^\dagger
\phi_{N,ij}
+
D_b^{(1)}
\left[a_i
\left(
\frac{2}{\mu}iD_0 d_{1,j}
\right)
\right]^\dagger
\phi_{N,ij}
+\cdots ,
\label{eq:Lweak_new}
\end{align}
where $\phi_{N,ij}$ is the source field for the $^{16}$N ground state, 
as a traceless symmetric tensor, 
and $a_i$ denotes the external axial-vector current.
The coefficients $C_a^{(1)}$ and $C_b^{(1)}$ represent the leading non-pole and pole couplings in
the $p$-wave, while $C_a^{(3)}$ and $C_b^{(3)}$ play the same role in the $f$-wave.
The coefficients $D_a^{(1)}$ and $D_b^{(1)}$ generate the first momentum-dependent correction in
the weak vertices for the $p$-wave channel. 
We note that the values of $C_b^{(1)}$ and $C_b^{(3)}$ are fixed by using the branching ratios 
of the $^{16}$N $\beta$-decay to the $1_1^-$ and $3_1^-$ states of $^{16}$O, respectively,
as functions of the effective range parameters~\cite{sa-epja21}. 

After integrating over the lepton phase space, the $\beta$-delayed $\alpha$ spectrum can be
written in the compact form
\begin{equation}
\frac{d\Gamma}{dE_\alpha}
=
\mathcal{N}\,
p\,
\mathcal{I}_\beta(E)
\left[
15\,C_\eta^2 W_1(\eta)\,|\mathcal{A}_1(E)|^2
+
\frac{28}{5}\,C_\eta^2 W_3(\eta)\,|\mathcal{A}_3(E)|^2
\right] ,
\label{eq:dGdEa_new}
\end{equation}
where
$E = (4/3)E_\alpha$, $p=\sqrt{2\mu E}$, and $\mathcal{I}_\beta(E)$ denotes the remaining leptonic phase-space integral, and $\mathcal{N}$ is an overall constant.
Using the notation of the present work,
$E_\nu = Q_m' - (E_e-m_e) - E$ and $Q_m' = 3.257~\mathrm{MeV}$,
so that~\cite{sa-epja21}
\begin{equation}
\mathcal{I}_\beta(E)
=
\int_0^{p_{e,\max}} dp_e
\int_{-1}^{1} dy\,
\frac{p_e^2 E_\nu}{2E_e}\,
F(Z,E_e)
\left(
E_\nu E_e 
\, -\frac{1}{3}\,\vec{p}_\nu \cdot \vec{p}_e
\right)\, ,
\label{eq:Ibeta_new}
\end{equation}
where $y=\cos\theta_{e\nu}$, with $\theta_{e\nu}$ denoting the angle
between the electron and neutrino momenta.
The maximum electron momentum is
$p_{e,\max}=\sqrt{\left(Q'_m+m_e-E\right)^2-m_e^2}.$

\begin{figure}
\centering
\includegraphics[width=0.35\textwidth]
{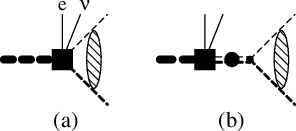}
\caption{
Feynman diagrams for $\beta$-delayed $\alpha$ distribution from $^{16}$N.
A very thick dashed line denotes the initial ground state of $^{16}$N, and filled boxes are
weak vertices coupled with the lepton current.
Diagram (a) represents the non-pole contribution, in which the final
two-body state is produced at the weak vertex, while diagram (b)
represents the pole contribution, in which the final two-body states
in the $l=1$ and $l=3$ channels are produced after propagation
through the dressed $^{16}$O states.
The remaining notation is the same as in Fig.~\ref{fig:16propagators}.
}
\label{fig:weak_amplitudes}
\end{figure}

The reduced amplitudes are calculated from the Feynman diagrams in
Fig.~\ref{fig:weak_amplitudes} and consist of two parts:
\begin{align}
\mathcal{A}_1(E)
&=
\mathcal{A}_{1,\mathrm{np}}(E)
+
\mathcal{A}_{1,\mathrm{pole}}(E)\,,
\label{eq:A1decomp_new}
\\[0.3em]
\mathcal{A}_3(E)
&=
\mathcal{A}_{3,\mathrm{np}}(E)
+
\mathcal{A}_{3,\mathrm{pole}}(E).
\label{eq:A3decomp_new}
\end{align}
Here the $p$-wave amplitude contains a non-pole part 
and a pole part proportional to the dressed $p$-wave propagator. 
At the schematic level,
\begin{align}
\mathcal{A}_{1,\mathrm{np}}(E)
&=
C_a^{(1)}
+
D_a^{(1)}\frac{p^2}{\mu^2}
+\cdots ,
\label{eq:A1np_new}
\\[0.3em]
\mathcal{A}_{1,\mathrm{pole}}(E)
&=
\left(
C_b^{(1)}
+
D_b^{(1)}\frac{p^2}{\mu^2}
+\cdots
\right)
\frac{1}{D_1(p)}~,
\label{eq:A1pole_new}
\\[0.3em]
\mathcal{A}_{3,\mathrm{np}}(E)
&=
C_a^{(3)}+\cdots ,
\\[0.3em]
\mathcal{A}_{3,\mathrm{pole}}(E)
&=
\left( C_b^{(3)} + \cdots \right)
\frac{1}{D_3(p)}\,.
\label{eq:A3pole_new}
\end{align}

The essential point is that the same inverse dressed propagators
$D_1(p)$ and $D_3(p)$, constrained by elastic
$\alpha$--$^{12}$C scattering, also enter the decay amplitude{\te s}.

This immediately explains the most important structural difference from the conventional
$R$-matrix interpretation.
In the cluster-EFT picture, the low-energy shoulder of the $^{16}$N spectrum is generated by the
interference between non-pole and pole contributions built on the same dressed propagator.
The secondary peak therefore does not require an explicit interference between independently
parametrized subthreshold and resonant levels.
This reorganization of the amplitude is a defining feature of the present approach and is the basis
for the joint analysis carried out in the following section.

\section{Data sets and fitting procedure}
\label{sec:datafit}

In this section, we describe the experimental data sets used in the analysis
and define the sector chi-square functions for the Azuma, Tang, and
elastic-scattering sectors. We then specify how these sector chi-squares are
combined in the independent fits and in the raw and sector-balanced
simultaneous fits. Finally, we summarize the DE searches used to locate 
the selected minima and the MCMC sampling used to determine parameter 
uncertainties and correlations.

\subsection{Experimental data sets}
\label{subsec:datasets}

The elastic data are those used in our global analysis of
$\alpha$-$^{12}$C scattering~\cite{mpha-prc26}.
They contain $N_{\rm el}=11\,392$ measurements over
$E_\alpha=2.6$--$6.7$~MeV and 32 laboratory angles from
$24.0^\circ$ to $165.9^\circ$.
The absolute differential cross section at $58.9^\circ$ is used as the
reference observable, while the measurements at the remaining angles enter
through ratios to this reference~\cite{tetal-prc09,mpha-prc26}.
Only the shared $p$-wave parameters $(r_1,P_1,Q_1)$ are varied in the present
joint analyses; all other elastic-sector parameters are fixed to the values
of Ref.~\cite{mpha-prc26}.
The baseline elastic likelihood uses the quoted pointwise uncertainties and
a diagonal covariance matrix.

The corrected Azuma spectrum contains
$N_{\rm A}=91$ points over
$E_\alpha=0.59$--$2.39$~MeV~\cite{Azuma1994},
whereas the Tang spectrum contains
$N_{\rm T}=93$ points over
$E_\alpha=0.45$--$2.31$~MeV~\cite{tang}.
Both data sets are tabulated with a nominal bin width of 20~keV.
For $E_\alpha>1.9$~MeV, we use the Tang yields and enlarged uncertainties
including the incomplete-stopping correction reported in
Ref.~\cite{tang}.
All published points and quoted uncertainties are retained; no
energy-dependent data selection is applied in the baseline fits.
No additional historical normalization factor is applied to the Tang data.
Instead, the overall coefficient $C_{\rm OAC}$ is fitted independently for
each BDAD spectrum.

We use the branching-ratio inputs reported by
Kirsebom \textit{et al.}~\cite{Kirsebom2018}: $b_{\beta,11}=(5.02\pm0.10)\times10^{-2}$, 
$b_{\beta,31} = 0.670 \pm 0.006$, 
$b_{\beta\alpha}=(1.59\pm0.06)\times10^{-5}$.
Here, $b_{\beta,11}$ and $b_{\beta,31}$ denote the $\beta$-feeding branching ratios to the subthreshold $1^-_1$ and $3^-_1$ states of $^{16}\mathrm{O}$, respectively, while $b_{\beta\alpha}$ denotes the total branching ratio for $\beta$-delayed $\alpha$ emission.
Their central values are treated as fixed external inputs in the present
analysis.
The quoted branching-ratio uncertainties are therefore not included in the
generalized-posterior intervals reported below.

\subsection{Sector chi-squares and analysis objectives}
\label{subsec:objectives}
\label{subsec:chisq}
For the Azuma, Tang, and elastic sectors,
$X\in\{A,T,{\rm el}\}$, we 
employ the standard chi-squared function
\begin{equation}
\chi_X^2(\boldsymbol{\theta})
=
\sum_{i=1}^{N_X}
\left[
\frac{
y_{X,i}^{\rm th}(\boldsymbol{\theta})
-y_{X,i}^{\rm exp}
}{
\sigma_{X,i}
}
\right]^2 .
\label{eq:chi_sector_final}
\end{equation}
Here $\boldsymbol{\theta}$ denotes the parameter vector appropriate to
the analysis under consideration.
For the independent fit to data set $d\in\{A,T\}$, we use the eight parameter vector, 
\begin{equation}
\boldsymbol{\theta}_{8,d}
=
\left(
C_{\rm OAC}^{d},
C_a^{(1)},
D_a^{(1)},
D_b^{(1)},
C_a^{(3)},
r_1,
P_1,
Q_1
\right).
\label{eq:theta_independent}
\end{equation}
For the simultaneous analyses, we use the nine-parameter vector,
\bea
\boldsymbol{\theta}_{\rm sim} = \left(C_{\rm OAC}^{A}, C_{\rm OAC}^{T}, C_a^{(1)},
D_a^{(1)}, D_b^{(1)}, C_a^{(3)}, r_1,P_1,Q_1 \right)\,,
\label{eq:theta_sim}
\eea
where the two overall normalizations are independent, 
whereas the weak-current and strong-sector parameters 
are common to the two spectra.

The sector chi-squares are constructed using the following error model.
Each BDAD datum is treated as an independent Gaussian measurement centered
on the theoretical prediction, with its quoted experimental uncertainty as
the standard deviation.
The normalized residuals of the elastic scattering data are 
also treated as Gaussian with the adopted
diagonal covariance matrix.
This error model is used whenever
$\chi_A^2$, $\chi_T^2$, or $\chi_{\rm el}^2$ is evaluated, both in the
DE searches and in the MCMC sampling.
It specifies the statistical model for the residuals; DE and MCMC are the
numerical procedures used to optimize and sample the resulting objectives.

For numerical convenience, the BDAD sector chi-squares are evaluated on the
transformed yield scale used in Ref.~\cite{sa-epja21}.
For data set $d\in\{A,T\}$, we define  
\begin{equation}
y_{d,i}^{\rm exp,tr}
=
{\cal F}_d(E_{\alpha,i})\,
y_{d,i}^{\rm exp,raw},
~~
\sigma_{d,i}^{\rm exp,tr}
=
{\cal F}_d(E_{\alpha,i})\,
\sigma_{d,i}^{\rm exp,raw},
~~
y_{d,i}^{\rm th,tr}(\boldsymbol{\theta})
=
{\cal F}_d(E_{\alpha,i})\,
y_{d,i}^{\rm th,raw}(\boldsymbol{\theta}),
\label{eq:bdad_transformation}
\end{equation}
where a common factor ${\cal F}_d(E_\alpha)$ and 
the transformed theoretical yeld $y_{d,i}^{th,tr}(\theta)$ are obtained from Eq.~(\ref{eq:dGdEa_new}) as
\bea
{\cal F}_d(E_\alpha)
 &=&
\frac{{\cal K}_d}
     {C_\eta^2\,p\,\mathcal{I}_\beta(E)}~ \,,
     \\
     y_{d,i}^{th,tr}(\theta) &=& C_{OAC}^d 
\left[
W_1(p_i)\,|\mathcal{A}_1(E)|^2 + \frac{28}{75} W_3(p_i)\,|\mathcal{A}_3(E)|^2
\right] \, ,
\eea
Here $\mathcal{K}_d$ is a fixed, 
data-set-dependent conversion constant and is not varied in the fit. 
No additional historical Azuma-to-Tang rescaling is applied. 
$C_{\mathrm{OAC}}^d$ is fitted independently for $d=A,T$.
Since the same transformation factor $\mathcal{F}_d$ is applied to the experimental yield, 
theoretical yield, and experimental uncertainty in Eq. (23), 
the pointwise pulls and hence $\chi^2_d$ are unchanged by the transformation.

The same sector chi-squares are used in all analyses.
The independent and simultaneous calculations differ only in how these
sector contributions are combined, because they address different questions.

\subsubsection{Independent fits.}
For an independent analysis of BDAD data set
$d\in\{A,T\}$ together with the elastic data, we minimize
\begin{equation}
 \mathcal{J}_{{\rm ind},d}
 =
 \chi_d^2
 +
 \frac{N_d}{N_{\rm el}}\chi_{\rm el}^2
 =
 N_d
 \left(
 \frac{\chi_d^2}{N_d}
 +
 \frac{\chi_{\rm el}^2}{N_{\rm el}}
 \right)\,,
 \label{eq:ind_balanced_objective}
\end{equation}
with the eight parameter vectors in Eq.~(\ref{eq:theta_independent}). 
The numerical elastic weights are 
$w_A = N_A/N_{el}  
= 7.9881 \times 10^{-3}$,
$w_T = N_T/N_{el} 
= 8.1636 \times 10^{-3}$.
This sector-balanced loss gives the BDAD and elastic contributions
comparable influence on the location of the optimum.
It is not the negative logarithm of the raw joint Gaussian likelihood.
Accordingly, intervals obtained from this objective are
generalized-posterior intervals conditional on the adopted weighting.

\subsubsection{Simultaneous fits.}
The primary compatibility test minimizes the raw joint objective
\begin{equation}
 \chi_{\rm sim,raw}^2
 =
 \chi_A^2+\chi_T^2+\chi_{\rm el}^2 \,,
 \label{eq:sim_raw_objective}
\end{equation}
with the nine-parameter vector in Eq.~(\ref{eq:theta_sim}).
Because no additional sector weight is introduced,
Eq.~(\ref{eq:sim_raw_objective}) is the raw joint chi-square associated
with the error model specified above.
Equivalently, up to an additive constant,
$-\chi_{\rm sim,raw}^2/2$ is the log-likelihood under the adopted
Gaussian error and covariance model.
The elastic prediction depends directly only on $(r_1,P_1,Q_1)$;
the compromise among the common weak-current coefficients is therefore
determined primarily by the two BDAD contributions.

As a sensitivity test, we also use the sector-balanced simultaneous objective
\begin{equation}
 \mathcal{J}_{\rm sim,bal}
 =
 \chi_A^2+\chi_T^2
 +
 \frac{N_A+N_T}{N_{\rm el}}\chi_{\rm el}^2 
 =
 (N_A+N_T)\left(
 \frac{\chi_A^2+\chi_T^2}{N_A+N_T}
 +
 \frac{\chi_{\rm el}^2}{N_{\rm el}}
 \right) \, ,
 \label{eq:sim_balanced_objective}
\end{equation}
where the weight is
$w_{sim} = (N_A + N_T) / N_{\rm el} 
= 1.6152 \times 10^{-2}$. 
This form gives the combined BDAD block and the elastic block comparable
influence.
It tests whether the conclusion from the raw fit is driven mainly by the much
larger number of elastic points.

The numerical values of
$\mathcal{J}_{{\rm ind},d}$,
$\chi_{\rm sim,raw}^2$, and
$\mathcal{J}_{\rm sim,bal}$ are not compared directly because they define
different estimators.
We compare instead the identically defined sector diagnostics
$\chi_A^2/N_A$, $\chi_T^2/N_T$, and
$\chi_{\rm el}^2/N_{\rm el}$.

\subsection{Optimization and uncertainty analysis}
\label{subsec:protocol}
The error model specified in Sec.~\ref{subsec:objectives} determines the
sector chi-squares, whereas DE and MCMC serve distinct numerical roles.
DE locates the minimum of the selected objective, and MCMC samples the
distribution constructed from that same objective.\footnote{
For detailed descriptions of the DE and affine-invariant
ensemble MCMC algorithms used in our cluster-EFT analysis, see
Secs.~III A and III B of Ref.~\cite{mpha-prc26}, respectively.
The present work follows the same general numerical framework,
while the objectives, parameter spaces, run settings, and convergence
criteria specific to this analysis are given here.}
For each analysis, the best-fit solution is first located with independent
DE searches using different random seeds and mutation
strategies~\cite{DE-paper,DE-book}. The lowest valid solution is retained only
after it is reproduced by at least one additional search within the prescribed
objective tolerance.  
The selected DE point defines the best-fit parameter vector used to
draw the central curves in Sec.~IV and to compute the corresponding
$\chi^2$ values.
The quoted parameter estimates, credible intervals, and uncertainty bands are 
obtained from the subsequent MCMC samples.

Let $\mathcal O(\boldsymbol\theta)$ denote the objective used in a given
analysis,
\begin{equation}
\mathcal O
=
\begin{cases}
\mathcal J_{{\rm ind},d},
& \text{independent Azuma or Tang fit},\\[1mm]
\chi^2_{\rm sim,raw},
& \text{raw simultaneous fit},\\[1mm]
\mathcal J_{\rm sim,bal},
& \text{balanced simultaneous fit}. \nonumber
\end{cases}
\label{eq:analysis_objective}
\end{equation}
Parameter correlations and conditional uncertainties are determined with
independent affine-invariant ensemble calculations using
\textsc{emcee}~\cite{emcee}.  The sampled density is
\begin{equation}
\ln p_{\mathcal O}(\boldsymbol\theta\mid{\rm data})
=
-\frac{1}{2}\mathcal O(\boldsymbol\theta)
+\ln\pi(\boldsymbol\theta)
+{\rm const.},
\label{eq:objective_posterior}
\end{equation}
where $\pi$ imposes the parameter bounds, the positive-residue condition, and
numerical-stability requirements.  For the raw simultaneous analysis,
Eq.~(\ref{eq:objective_posterior}) is the Gaussian posterior under the adopted
pointwise error model and diagonal elastic covariance.  For the independent
and balanced analyses, it defines a generalized posterior conditional on the
corresponding sector weights.

Three independent ensemble chains are used for each reported posterior.
Sampling is accepted when every parameter satisfies
$\max\widehat R\leq1.01$,
$N_{\rm eff}\geq3000$,
$\frac{N_{\rm post}}{\max\tau}\geq50$,
with an ensemble-averaged acceptance fraction between 0.15 and 0.60.\footnote{
The extrema are taken over all fitted parameters. The split-$\widehat{R}$ statistic 
compares between-chain and within-chain variation; values near unity indicate 
agreement among independently initialized chains, and $\widehat{R}\leq1.01$ 
is a stringent modern criterion. $N_{\mathrm{eff}}$ is the number of effectively 
independent draws after accounting for autocorrelation; the requirement 
$N_{\mathrm{eff}}\geq3000$ is a conservative analysis-specific choice for 
stable posterior quantiles. The integrated autocorrelation time $\tau$ 
measures the correlation length of the chain, so $N_{\mathrm{post}}/\max\tau\geq50$ 
requires the retained chain to span at least 50 autocorrelation times of the 
slowest-mixing parameter, consistent with practical \texttt{emcee} guidance. 
The acceptance-fraction interval is an additional empirical mixing check. 
These diagnostics establish sampling convergence within the selected posterior mode, 
not global-mode completeness or model adequacy.}
Equal numbers of retained samples from the independent chains are then pooled.
For each parameter, 
we report the posterior median and the central 68\% credible interval, 
defined by the 16th and 84th percentiles.
The uncertainty bands are the corresponding pointwise percentiles of the
propagated theoretical curves and do not include an additional draw of
experimental observation noise. 

\section{Independent and simultaneous joint-fit results}
\label{sec:results}
We first perform two independent joint analyses: Azuma plus elastic
scattering and Tang plus elastic scattering.
Both analyses use the same elastic data set, cluster-EFT amplitude{\te s},
branching-ratio inputs, parameter ranges, and numerical procedure.
We then test the mutual compatibility of the two BDAD spectra by fitting
Azuma, Tang, and elastic scattering simultaneously, using both raw and
sector-balanced objectives. 
The central curves of the observables correspond to the lowest DE solutions, while parameter
intervals are determined from the pooled MCMC samples described in
Sec.~\ref{subsec:protocol}.

\subsection{Fit quality and numerical stability}
\label{subsec:numerical_quality}
Table~\ref{tab:fit_quality_final8d} summarizes the fit quality and numerical
diagnostics of the two independent 8D analyses.~\footnote{
In Table~\ref{tab:fit_quality_final8d}, ``DE recovery'' is reported as $n_{\rm rec}/n_{\rm run}$, 
where $n_{\rm rec}$ is the number of independent DE searches that 
reproduced the lowest identified minimum within the adopted objective tolerance, 
and $n_{\rm run}$ is the total number of searches.
}
Relative to the previous fixed-propagator calculation
of Ref.~\cite{sa-epja21}, the BDAD $\chi^2/N$ decreases from
4.06 to 1.73 for  the Azuma fit and from 3.56 to 2.03 for the Tang fit.
These correspond to improvements of approximately 57\% and 43\%,
respectively.

\begin{table}
\caption{
Fit quality and numerical diagnostics for the independent 8D analyses.
The previously reported BDAD values are taken from Ref.~\cite{sa-epja21}.
}
\label{tab:fit_quality_final8d}
\begin{ruledtabular}
\begin{tabular}{lcc}
Quantity
& Azuma
& Tang
\\
\hline
$N_{\rm BDAD}$
& 91
& 93
\\
Previous $\chi^2_{\rm BDAD}/N_{\rm BDAD}$
& 4.06
& 3.56
\\
Present $\chi^2_{\rm BDAD}/N_{\rm BDAD}$
& 1.73
& 2.03
\\
$\chi^2_{\rm el}/N_{\rm el}$
& 6.42
& 6.70
\\
DE recovery
& $3/3$
& $2/3$
\\
Maximum $\widehat R$
& 1.007
& 1.009
\\
Minimum ESS
& $1.04\times10^5$
& $8.87\times10^4$
\\
\end{tabular}
\end{ruledtabular}
\end{table}
All three Azuma DE searches converge to the same minimum, with 
a spread below $4\times10^{-9}$ in
$\mathcal J_{{\rm ind},A}$.
For Tang, two searches recover the lowest identified minimum, while the
third converges to a nearby solution with
$\Delta\mathcal J_{{\rm ind},T}=0.400$.
The two Tang solutions have nearly identical strong-sector parameters and
differ mainly in the weak-current coefficients, including the sign of
$C_a^{(3)}$.
The results presented below correspond to the lowest solution.
The nearby second Tang solution 
and the conditional nature of the reported posterior
are discussed in Appendix~\ref{app:robustness}.
The ensemble calculations show stable sampling around the selected minima.
For both datasets, the maximum split-$\widehat R$ remains below $1.01$, and
the minimum pooled effective sample size exceeds $8.87 \times 10^4$.
These diagnostics establish numerical convergence within the sampled
posterior modes.

\subsection{BDAD spectra and line-shape comparison}
\label{subsec:bdad_results}

\begin{figure*}
\centering
\includegraphics[width=0.82\textwidth]
{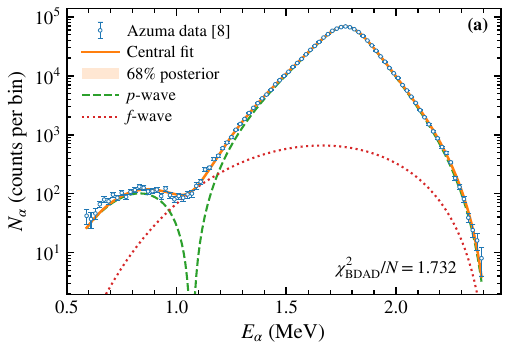}
\includegraphics[width=0.82\textwidth]
{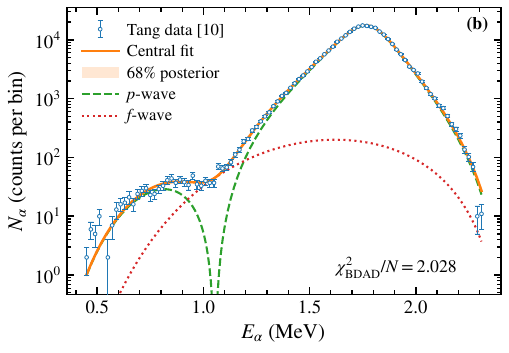}
\caption{
Independent eight-parameter joint fits to the
Azuma (a) and Tang (b)
$^{16}$N $\beta$-delayed $\alpha$ spectra.
The solid curves show the total spectra evaluated at the lowest DE solutions,
while the shaded bands denote the pointwise 68\% generalized-posterior
intervals obtained from the pooled MCMC samples.
The dashed and dotted curves show the corresponding central $p$- and
$f$-wave contributions, respectively.
All published data points and quoted uncertainties are included in the fits.
}
\label{fig:bdad_final8d}
\end{figure*}

Figure~\ref{fig:bdad_final8d} compares the two BDAD spectra with the
central DE curves and pointwise 68\% generalized-posterior intervals.
Both fits reproduce the low-energy shoulder, the minimum near
$E_\alpha\simeq1.0$~MeV, the broad maximum, and the subsequent
high-energy falloff.

The broad maximum is dominated by the $p$-wave contribution in both analyses.
The $f$-wave becomes relatively important near the $p$-wave minimum and
therefore affects the depth and shape of the low-energy structure.
This common decomposition shows that the main continuum peak is governed
primarily by the shared $p$-wave dynamics, whereas the low-energy region
provides greater sensitivity to the  weak-decay amplitudes.

For the Azuma sector, 
the largest absolute pull in the $\chi^2/N$ calculation is 4.21 at
$E_\alpha=2.01$~MeV, while 
the remaining residuals are distributed over several bins.
For the Tang sector, the larger $\chi^2/N$ is driven mainly by the two
highest-energy points at 2.29 and 2.31~MeV.
The visually isolated point at 0.55~MeV has comparatively little statistical
leverage.
Both datasets also exhibit a localized discrepancy near
$E_\alpha\simeq2.01$~MeV.
It might be a remnant effect of the sharp resonant $2_2^+$ state of $^{16}$O at 
$E_\alpha \simeq 2.012$~MeV
($E_x = 9.845$~MeV).

\begin{figure*}[t]
\centering
\includegraphics[width=0.92\textwidth]{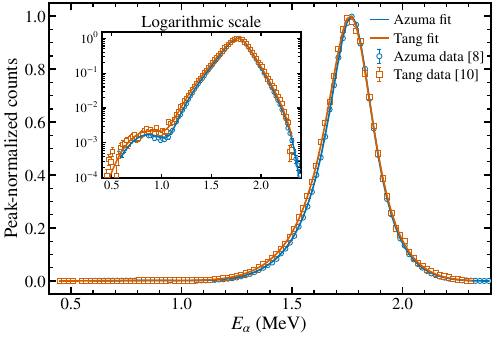}
\caption{
Peak-normalized comparison of the Azuma and Tang spectra and their
8D fits.
Each dataset and curve is divided by the maximum of its own central fit in
$1.60\leq E_\alpha\leq1.90$~MeV.
The logarithmic inset emphasizes the low-energy structure and the
high-energy endpoint.
The rescaling is used only for comparison and does not enter the fits.
}
\label{fig:bdad_normalized}
\end{figure*}

To compare the line shapes independently of their absolute normalizations,
Fig.~\ref{fig:bdad_normalized} scales each spectrum and central curve to
the fitted maximum in
$1.60\leq E_\alpha\leq1.90$~MeV.
After this rescaling, the positions and widths of the broad maxima are nearly
identical.
The remaining differences are concentrated in the low-energy shoulder,
the rising side of the peak, and the Tang endpoint.
This normalization is used only for visualization and does not enter the
fit.

\subsection{Parameter estimates and strong-sector stability}
\label{subsec:parameter_results}

The present posterior estimates are compared with the earlier
fixed-propagator results in Table~\ref{tab:weak_parameter_comparison}.
The reported values are posterior medians with 16th--84th percentile
intervals.
Only coefficients with compatible conventions are compared directly with
Ref.~\cite{sa-epja21}. 
For the Azuma sector, the magnitudes of
$C_{\rm OAC}$, $C_a^{(1)}$, and $D_a^{(1)}$ are reduced relative to the
previous analysis by approximately 10\%, 17\%, and 24\%, respectively.
For the Tang sector, $D_a^{(1)}$ remains nearly unchanged, whereas
$C_{\rm OAC}$ and $C_a^{(1)}$ decrease by approximately 44\% and 15\%.
The improved fits therefore retain a measurable dataset dependence in the
normalization and weak-current amplitudes.

\begin{table*}[t]
\caption{
Weak-sector parameters from the previous fixed-propagator analysis
of Ref.~\cite{sa-epja21} and the present 8D joint fits.
The present values are generalized-posterior medians with 68\% intervals.
The previous definitions of $D_b^{(1)}$ and the $f$-wave coefficient are not
directly equivalent to the present conventions and are therefore not compared.
}
\label{tab:weak_parameter_comparison}
\begin{ruledtabular}
\begin{tabular}{lcccc}
Quantity
& \shortstack{Previous\\ Azuma}
& \shortstack{Present\\ Azuma}
& \shortstack{Previous\\ Tang}
& \shortstack{Present\\ Tang} \\
\hline
$C_{\rm OAC}\ (10^6\,{\rm MeV}^{-6})$
& $7.2(4)$
& $6.50^{+0.58}_{-0.55}$
& $4.22(7)$
& $2.38^{+0.35}_{-0.32}$
\\
$C_a^{(1)}\ (10^{-3}\,{\rm MeV}^{-2})$
& $-6.9(2)$
& $-5.73^{+0.43}_{-0.40}$
& $-9.46(5)$
& $-8.00^{+0.68}_{-0.59}$
\\
$D_a^{(1)}\ ({\rm MeV}^{-2})$
& $2.61(9)$
& $2.00^{+0.16}_{-0.18}$
& $3.36(3)$
& $3.34^{+0.25}_{-0.29}$
\\
$D_b^{(1)}\ (10^3\,{\rm MeV})$
& ---
& $-8.88^{+0.09}_{-0.10}$
& ---
& $-8.01^{+0.12}_{-0.14}$
\\
$C_a^{(3)}\ (10^{-7}\,{\rm MeV}^{-4})$
& ---
& $-2.089^{+0.069}_{-0.075}$
& ---
& $+0.658^{+0.130}_{-0.114}$
\\
\end{tabular}
\end{ruledtabular}
\end{table*}

The shared effective-range parameters and selected derived quantities are
listed in Table~\ref{tab:strong_diagnostics_8d}.
The effective-range parameters remain close to the elastic reference values in Ref.~\cite{sa-epja21},
despite the substantial improvement of the BDAD fits. 
The Azuma solution changes each of $(r_1,P_1,Q_1)$ by less than 0.8\% relative to the reference solution.
The Tang-conditioned result shows a somewhat larger but still modest change,
with the largest shift, approximately 3.6\%, occurring in $Q_1$.
Thus, the improved BDAD description does not require a qualitatively
different $p$-wave strong interaction.

The $p$-wave ANC shows moderate dataset dependence, whereas the
$f$-wave ANC is unchanged because the $l=3$ strong sector is fixed.
Because the effective-range and 
weak-current parameters are strongly correlated,
these central values should be regarded as diagnostics rather than
independent determinations. 
The pooled posterior structures and representative parameter correlations
are presented in Appendix~\ref{app:robustness}.

\begin{table*}[t]
\caption{
Shared $p$-wave effective-range parameters and derived diagnostics.
The first three rows give posterior medians and 68\% intervals.
The remaining quantities are evaluated at the central DE solutions.
The reference values are taken from Ref.~\cite{sa-epja21}.
}
\label{tab:strong_diagnostics_8d}
\begin{ruledtabular}
\begin{tabular}{lccc}
Quantity
& Reference
& Azuma 8D
& Tang 8D
\\
\hline
$r_1\ ({\rm fm}^{-1})$
& $0.415273(9)$
& $0.415230(21)$
& $0.414996(28)$
\\
$P_1\ ({\rm fm})$
& $-0.57473(9)$
& $-0.57523(24)$
& $-0.57740(29)$
\\
$Q_1\ ({\rm fm}^{3})$
& $0.02018(3)$
& $0.020027(82)$
& $0.019455(94)$
\\
$|C_{b,1^-_1}|\ ({\rm fm}^{-1/2})$
& $1.832\times10^{14}$
& $1.810\times10^{14}$
& $1.942\times10^{14}$
\\
$|C_{b,3^-_1}|\ ({\rm fm}^{-1/2})$
& $2.4(4)\times10^2$
& $2.252\times10^2$
& $2.252\times10^2$
\end{tabular}
\end{ruledtabular}
\end{table*}

\subsection{Elastic-sector consistency}
\label{subsec:elastic_results}

Only the shared $p$-wave effective-range parameters
$(r_1,P_1,Q_1)$ are varied in the present joint analyses.
All remaining elastic-sector parameters are fixed to the global solution
of Ref.~\cite{mpha-prc26}.

\begin{figure*}
\centering
\includegraphics[width=0.90\textwidth]{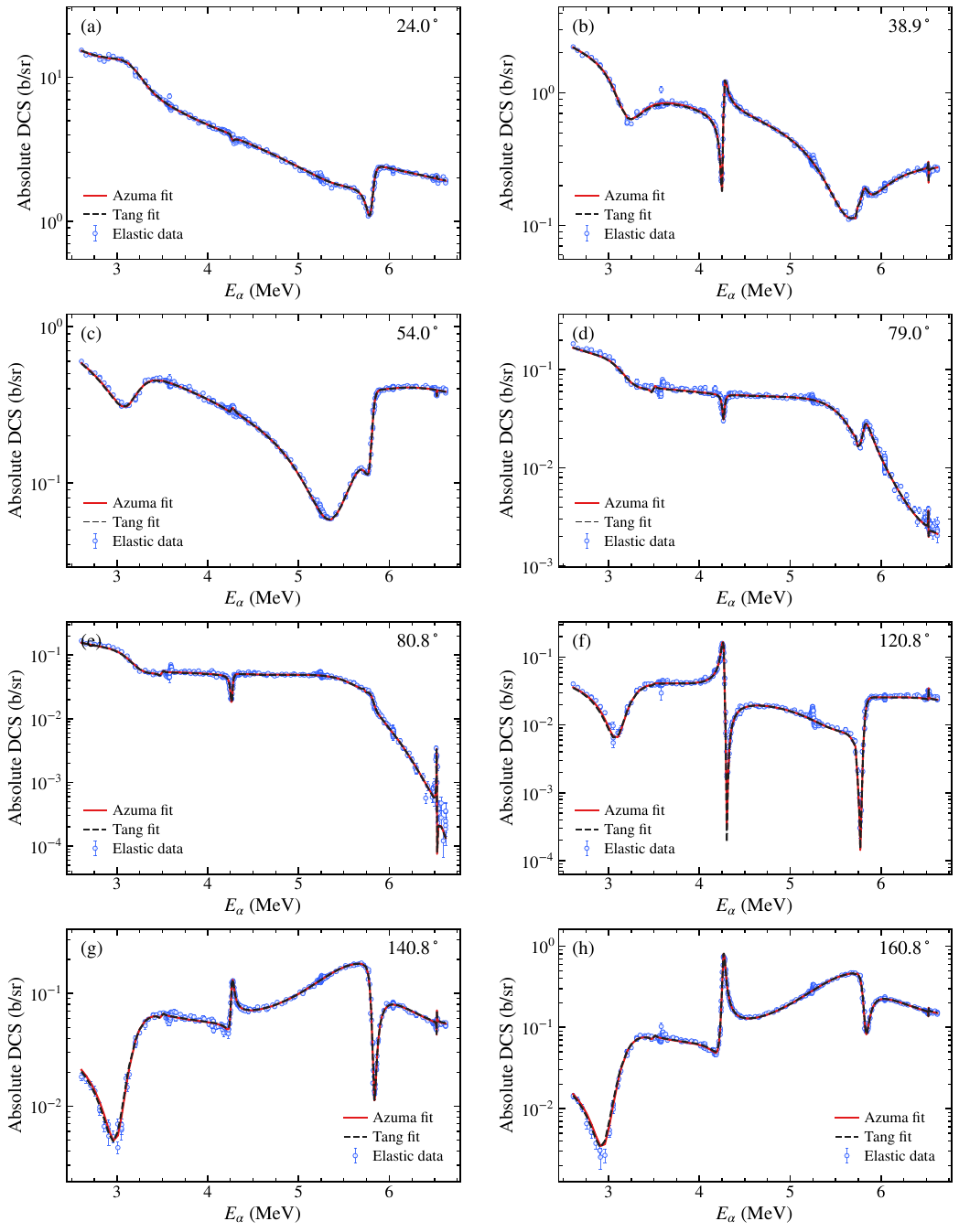}
\caption{
Representative absolute differential cross sections for elastic
$\alpha$--$^{12}$C scattering.
Only the shared $p$-wave effective-range parameters
$(r_1,P_1,Q_1)$ are varied in the present joint fits; all remaining
elastic parameters are fixed to the global solution of
Ref.~\cite{mpha-prc26}.
The curves therefore show the effect of conditioning the shared
$p$-wave propagator on the Azuma or Tang BDAD spectrum and provide a
consistency check against the full elastic data, rather than a new global
fit of the complete elastic amplitude.
}
\label{fig:elastic_final8d_comparison}
\end{figure*}

Figure~\ref{fig:elastic_final8d_comparison} therefore does not represent
a new global refit of the complete elastic amplitude.
Rather, it tests whether the BDAD-conditioned changes of the shared
$p$-wave propagator remain compatible with the full elastic
differential-cross-section data when combined with the fixed nonshared
elastic parameters.
The two calculations of the Azuma and Tang sectors 
are nearly indistinguishable over the measured energy
and angular ranges and retain the global description obtained in the
elastic-only analysis.
Their fit qualities are 
$\chi^2_{\rm el}/N_{\rm el} =$ 6.4194 for the Azuma fit and 6.696 for 
the Tang fit, respectively.
Thus, the substantial improvement of the BDAD spectra is achieved without
a qualitative distortion of the common elastic amplitude.

The elastic $\chi^2$ is dominated by a small number of narrow energy regions:
approximately 22\% of the data account for 67\% and 64\% of
$\chi^2_{\rm el}$ in the Azuma- and Tang-conditioned fits,
respectively.
The $\chi^2/N$ values therefore reflect localized discrepancies
rather than a broad failure of the elastic line shape.
Because the present likelihood assumes a diagonal covariance matrix,
the quoted elastic intervals remain conditional on the treatment of
energy resolution, angular averaging, calibration, and correlations
induced by the common $58.9^\circ$ reference measurement.

\subsection{Simultaneous common-current compatibility tests}
\label{subsec:simultaneous_fit}
We next test whether the Azuma and Tang spectra can be described by one common
set of strong and weak-current coefficients, while retaining independent
overall normalizations. The nine-parameter vector is defined in Eq.~(\ref{eq:theta_sim}), 
and the raw and sector-balanced objectives
are given in Eqs.~(\ref{eq:sim_raw_objective}) and
(\ref{eq:sim_balanced_objective}), respectively.  The raw objective retains
the pointwise Gaussian weighting of all reported data, whereas the balanced
objective tests sensitivity to the much larger elastic sample.
The sector chi-squares are evaluated identically in all analyses; only their
weights in the optimization objective differ.  Consequently,
$\chi_A^2/N_A$, $\chi_T^2/N_T$, and
$\chi_{\rm el}^2/N_{\rm el}$ can be compared directly across fits, whereas
the absolute raw and balanced objective values cannot.

The lowest raw DE solution is reproduced by an independent search within
$2.3\times10^{-7}$ in $\chi^2_{\rm sim,raw}$.  Its pooled chains give
$\max\widehat R=1.0033$ and a minimum effective sample size of
$1.99\times10^5$.  For the balanced analysis, the selected minimum is
reproduced by two of four DE searches within $3.1\times10^{-9}$ in
$\mathcal J_{\rm sim,bal}$; the pooled chains give
$\max\widehat R=1.0082$ and a minimum effective sample size of
$3.42\times10^4$.  Both analyses satisfy the adopted convergence criteria.
Linear correlations of the fitted parameters,
selected Pearson correlation coefficients, 
and numerical convergence diagnostics on simultaneous fits are discussed in detail
in Appendix~\ref{app:robustness}. 

The raw and sector-balanced analyses therefore lead to the same qualitative
compatibility conclusion.  Independent normalizations remove the scale
difference and the common strong propagator remains stable, but neither
objective reproduces both BDAD spectra with the quality obtained in the
independent fits.  This is a model-compatibility statement, not evidence that
the underlying physical weak amplitudes are intrinsically different.  The
objective dependence of the fitted weak-current coefficients further shows
that the simultaneous calculation should not be interpreted as a unique
extraction of the decay amplitude.

\begin{figure*}[t]
\centering
\includegraphics[width=0.90\textwidth]{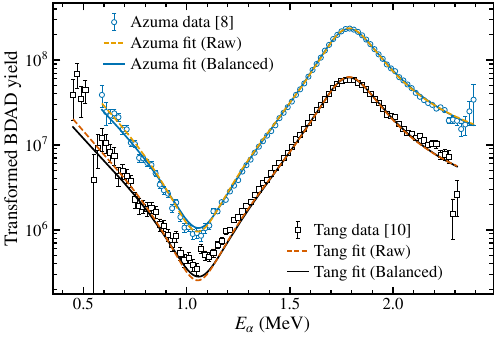}
\caption{
Raw and sector-balanced simultaneous common-current 
fits to the Azuma and Tang
$^{16}$N $\beta$-delayed $\alpha$ spectra.  
The data and predictions are shown on the transformed yield 
introduced in Eq.~(\ref{eq:bdad_transformation}).
Each set of the two curves uses common strong and weak-current coefficients and 
independently fitted overall normalizations.  
}
\label{fig:bdad_simultaneous_raw}
\end{figure*}
In Fig.~\ref{fig:bdad_simultaneous_raw}, 
we plot the transformed BDAD yields introduced 
in Eq.~(\ref{eq:bdad_transformation}) as functions of $E_\alpha$. 
The experimental data points and errors are also transformed and included in the figure.
Four lines are plotted in the figure by using the two sets of fitted values of the parameters in the raw and 
sector-balanced analysis. The difference between the lines of the Azuma and Tang sectors is 
due to the fitted overall constant. 
One can see a difficulty in fitting the two data sets simultaneously using an overall constant alone. 
In addition, 
the difference between the two objectives is largest on the
low-energy side of the spectra,
$0.5\lesssim E_\alpha\lesssim1.0$~MeV.
This objective dependence should be propagated in a future
extrapolation of the 
$S$ factor of the $E1$ transition of the radiative capture to the Gamow region.

\begin{table*}[t]
\caption{
Sector-wise fit diagnostics for the independent and simultaneous analyses.
All entries are unweighted sector quantities and are therefore directly
comparable across rows.  The absolute raw and balanced objectives are not
compared because they define different estimators.
}
\label{tab:simultaneous_fit_quality}
\begin{ruledtabular}
\begin{tabular}{lccccc}
Analysis
& Objective
& $\chi_A^2/N_A$
& $\chi_T^2/N_T$
& $\chi_{\rm el}^2/N_{\rm el}$
& $(\chi_A^2+\chi_T^2)/(N_A+N_T)$
\\
\hline
Azuma-conditioned 8D
& $\mathcal J_{{\rm ind},A}$
& 1.7317
& ---
& 6.4194
& ---
\\
Tang-conditioned 8D
& $\mathcal J_{{\rm ind},T}$
& ---
& 2.0285
& 6.6962
& ---
\\
Simultaneous common current 
& $\chi^2_{\rm sim,raw}$
& 4.3194
& 10.9011
& 6.3663
& 7.6460
\\
Simultaneous common current 
& $\mathcal J_{\rm sim,bal}$
& 3.3082
& 8.9926
& 6.4662
& 6.1813
\\
\end{tabular}
\end{ruledtabular}
\end{table*}

As displayed in Table \ref{tab:simultaneous_fit_quality},
the sector-balanced fit improves the agreement with both BDAD data sets relative to the raw fit: $\chi_A^2/N_A$ decreases from 4.3194 to 3.3082 for the Azuma sector,
and $\chi_T^2/N_T$ decreases from 10.9011 to 8.9926 for the Tang sector, while the
combined BDAD value decreases from 7.6460 to 6.1813. 
The elastic contribution changes only from 6.3663 to 6.4662. Thus, reducing the relative elastic weight improves the compromise between the two decay spectra but does not recover the quality of the independent fits. The remaining common-current 
tension is therefore not caused solely by the much larger number of elastic
data points.

The independent normalization coefficients remove the overall scale
difference between the two measurements.  Their fitted ratio is nearly
objective independent,
\begin{equation}
\frac{C_{\rm OAC}^{A}}{C_{\rm OAC}^{T}}
=
3.7489\quad\text{(raw)},
\qquad
3.7425\quad\text{(balanced)}. \nonumber
\label{eq:raw_balanced_normalization_ratio}
\end{equation}
The deterioration of the simultaneous fits therefore reflects a line-shape
compromise rather than a simple normalization offset.  In both analyses, the
residual extends over several adjacent bins on the rising flank and broad
maximum and is not confined to the Tang endpoint points.

The common weak-current coefficients are more sensitive to the objective.
Most notably, the fitted $f$-wave coefficient changes from
$C_a^{(3)} = -0.17696\times10^{-6}\,\mathrm{MeV}^{-4}$ to $+0.05006\times10^{-6}\,\mathrm{MeV}^{-4}$
when the raw objective is replaced by the balanced one.  The line-shape
incompatibility is therefore robust, but the detailed common weak-current
parameters should not be interpreted as an objective-independent physical
extraction.

The shared strong-sector parameters also remain stable under the change in
the objective function, with
$(r_1,P_1,Q_1)_{\rm raw}=(0.4153406,-0.5742015,0.0202981)$ and
$(r_1,P_1,Q_1)_{\rm bal}=(0.4151760,-0.5757270,0.0198962)$,
where the units of $r_1$, $P_1$, and $Q_1$ are
$\mathrm{fm}^{-1}$, $\mathrm{fm}$, and $\mathrm{fm}^{3}$, respectively.
The relative changes
are approximately 0.04\%, 0.27\%, and 2.0\%, confirming that the common
$p$-wave propagator is insensitive to the relative weighting of the elastic
and BDAD sectors. 

\section{Discussion}
\label{sec:discussion}
The independent analyses show that the Azuma and Tang spectra are separately
compatible with nearly the same elastic-scattering-constrained $p$-wave
continuum.  The raw and sector-balanced simultaneous fits sharpen this
conclusion.  Reducing the elastic weight lowers the combined BDAD value from
7.6460 to 6.1813, but the Azuma and Tang contributions remain well above the
values obtained in their independent fits.  The common-current
tension is
therefore not an artifact of the much larger elastic data set.
The strong-sector parameters and the Azuma-to-Tang normalization ratio are
stable under the change of objective.  In contrast, the common weak-current
coefficients change appreciably, including a sign change of $C_a^{(3)}$.
Thus, the robust result is the inability of the minimal common-current model to
reproduce both line shapes; the detailed common-current parameters are
objective dependent and should not be assigned a unique physical
interpretation.  This remains a model-compatibility statement and should not
be taken as direct evidence that the underlying physical weak amplitudes are
intrinsically different.

We retain the 8D amplitude as the baseline because it provides the minimal 
and most transparent connection between elastic scattering and weak decay. 
Higher-dimensional extensions can improve the individual BDAD fits, 
but their additional smooth coefficients are not uniquely associated with 
physical ($^{16}\mathrm{O}$) levels and are therefore treated as parametrization dependence. 
The fitted parameters are strongly correlated, so the interval quoted
for one parameter should not be interpreted independently of the others.
The uncertainty of a derived quantity should therefore be evaluated
using the full parameter samples generated by the MCMC calculation,
rather than by varying one parameter at a time. In addition, the Tang
fit has a nearby second solution with the opposite sign of
$C_a^{(3)}$; this discrete alternative should be examined separately
whenever it produces a non-negligible change in the quantity of interest,
as discussed in Appendix~\ref{app:robustness}.

The next step is a response-aware simultaneous analysis including
experiment-specific energy calibration, resolution, and response functions. 
Such a calculation will determine whether the remaining Azuma--Tang
difference is experimental or requires an extension of the weak amplitude.
A fully unified cluster-EFT analysis of elastic scattering, both BDAD spectra,
and $^{12}$C$(\alpha,\gamma)^{16}$O is currently in progress.
Its purpose is to propagate the correlated strong-, weak-, and
electromagnetic-sector uncertainties consistently to the extrapolated
$E1$ strength in the helium-burning Gamow window.

\section{Summary and outlook}
\label{sec:summary}
We have analyzed the $^{16}$N $\beta$-delayed $\alpha$ spectra of
Azuma \textit{et al.} and Tang \textit{et al.} within cluster EFT, using the
same elastic $\alpha$--$^{12}$C data and a common dressed $p$-wave
$^{16}$O propagator. 
Independent eight-dimensional (8D) fits yield
$\chi_{\rm BDAD}^2/N_{\rm BDAD} = 1.732$ for Azuma and $2.028$ for Tang,
compared with the previous fixed-propagator values of $4.06$ and $3.56$,
respectively.
These improvements are obtained without an additional effective $1^-$ resonance or
phenomenological smooth remainder, and the fitted effective-range parameters
remain close to the elastic reference solution.

The simultaneous nine-parameter analysis imposes common strong and
weak-current coefficients on both spectra while retaining independent overall
normalizations.  The sector-wise fit qualities are
\begin{equation}
\left(
\frac{\chi_A^2}{N_A},
\frac{\chi_T^2}{N_T},
\frac{\chi_{\rm el}^2}{N_{\rm el}}
\right)
=
\begin{cases}
(4.319,10.901,6.366), & \text{raw},\\
(3.308,8.993,6.466), & \text{balanced}. \nonumber
\end{cases}
\label{eq:summary_simultaneous_results}
\end{equation}
The balanced objective improves the decay sectors but does not recover the
quality of the independent fits.  The common-current tension is therefore not
generated by the numerical size of the elastic sample.

The fitted normalization ratio is nearly unchanged,
$C_{\rm OAC}^{A}/C_{\rm OAC}^{T}=3.7489$ and 3.7425 for the raw and balanced
fits, respectively, and the shared strong-sector parameters are likewise
stable.  The common weak-current coefficients are more sensitive to the
objective, with $C_a^{(3)}$ changing sign.  
The robust conclusion is therefore that the two spectra are separately
compatible with a common strong continuum, but the minimal common
weak-current amplitude does not describe them simultaneously and its
fitted coefficients depend on the objective.
This conclusion is conditional on the present response-free common-{\te current} 
model and does not establish an intrinsic difference between the physical
decay amplitudes.
The 8D amplitude is retained as the baseline because it provides the most
economical connection between elastic scattering and weak decay.
Higher-dimensional variants quantify residual parametrization dependence, but
their additional smooth coefficients are not identified with physical
$^{16}$O levels.  The quoted intervals also remain conditional on the adopted
elastic covariance, fixed branching-ratio inputs, and selected amplitude
branches.

We do not extract a new value of $S_{E1}(300\,\mathrm{keV})$ in the present work.
Radiative capture contains electromagnetic short-distance couplings that are
not fixed by elastic scattering or BDAD alone.  A response-aware combined
analysis of elastic scattering, both BDAD spectra, and
$^{12}$C$(\alpha,\gamma)^{16}$O is in progress~\cite{UnifiedCaptureWIP}.
It will propagate the correlated strong, weak, electromagnetic, experimental,
and EFT uncertainties to the extrapolated $E1$ strength in the
helium-burning Gamow window.

\section*{Acknowledgments}
J.P. was supported under Grants No.~RS-2021-NR060129,
No.~RS-2022-NR070836, No.~RS-2025-24533596, and
No.~RS-2025-25400847, funded by the Korean government through the
Ministry of Science and ICT and the Ministry of Education.
M.-H.M. was supported by the National Research Foundation of Korea (NRF) grant funded by the Korean government (MSIT) under Grant Nos.~RS-2026-25487837 and RS-2018-NR031074.
S.-I.A. was supported under Grant No.~RS-2025-16065411.

\appendix
\clearpage
\section{Composite $^{16}$O states in the elastic scattering amplitudes}
\label{app:state_content}

The index $i$ in the composite field $d_{(li)}$ labels the ordering of
states with the same orbital angular momentum $l$ 
(and is not an EFT expansion order).
The state content of the elastic {\te scattering } 
amplitude is summarized in
Table~\ref{tab:app_state_content}.
The broad $1^-_2$ and $3^-_2$ resonances are absorbed into the same
effective-range expansion (ERE) propagators for the subthreshold $1^-_1$ and $3^-_1$
states. 
Only the $l=1$ effective-range parameters are varied in the 
{\te present study.}
At each sampled point, 
the $1^-_1$ pole weak coefficient, $C_b^{(1)}$, is recalculated from
the current $p$-wave residue and the fixed branching-ratio input.
The $l=3$ effective-range parameters 
and the $3^-_1$ pole weak coefficient, $C_b^{(3)}$, 
are kept at their reference values.
All other elastic parameters remain fixed to the solution of
Ref.~\cite{mpha-prc26}.

\begin{table*}[t]
\caption{
$^{16}$O states represented in the elastic
$\alpha$-$^{12}$C scattering amplitudes and their role in the present joint analysis.
The assignments follow Refs.~\cite{sa-prc23,mpha-prc26}.
}
\label{tab:app_state_content}
\begin{ruledtabular}
\begin{tabular}{cccc}
Channel
& States in the ERE propagator
& Additional resonances
& Treatment in the present fit
\\
\hline
$0^+$
& $0_1^+,\,0_2^+$
& $0_3^+,\,0_4^+$
& Fixed elastic sector
\\
$1^-$
& $1_1^-,\,1_2^-$
& $1_3^-$
& $(r_1,P_1,Q_1)$ varied
\\
$2^+$
& $2_1^+$
& $2_2^+,\,2_3^+,\,2_4^+$
& Fixed elastic sector
\\
$3^-$
& $3_1^-,\,3_2^-$
& $3_3^-$
& Strong sector fixed; $C_a^{(3)}$ varied
\\
$4^+$
& ---
& $4_1^+,\,4_2^+,\,4_3^+$
& Fixed elastic sector
\\
$5^-$
& ---
& $5_1^-$
& Fixed elastic sector
\\
$6^+$
& ---
& $6_1^+$
& Fixed elastic sector
\\
\end{tabular}
\end{ruledtabular}
\end{table*}

\section{Posterior structure and numerical diagnostics}
\label{app:robustness}
This appendix summarizes the posterior structure and numerical convergence of
the independent and simultaneous analyses.
For the independent Azuma- and Tang-conditioned fits, the maximum
split-$\widehat R$ values are 1.007 and 1.009, and the minimum pooled
effective sample sizes are
$1.0\times10^5$ and $8.9\times10^4$, respectively.
The corresponding diagnostics for the simultaneous fits are given below.
All reported posteriors satisfy the adopted convergence criteria within their
selected modes.

\begin{figure*}[p]
\centering
\includegraphics[width=0.98\textwidth]{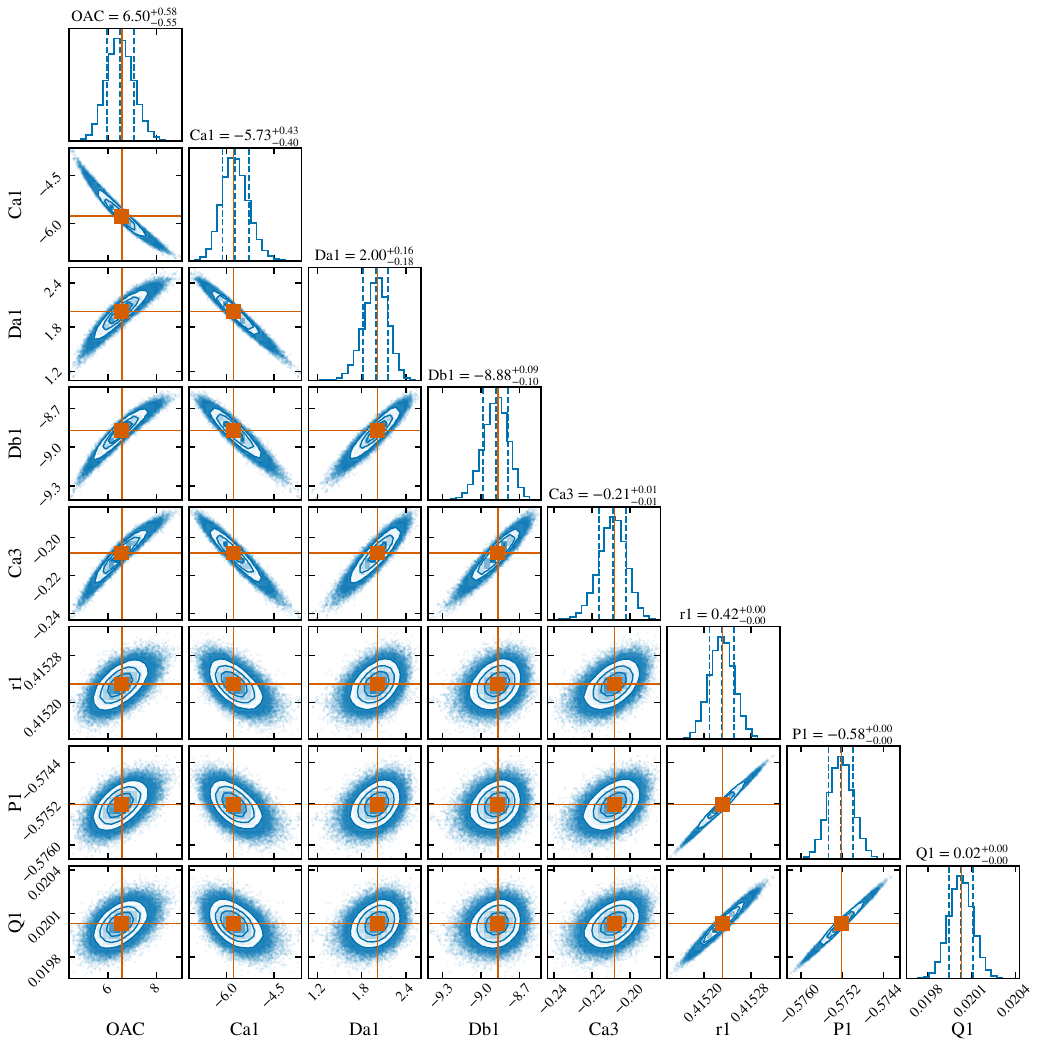}
\caption{
Pooled posterior distributions for the independent Azuma-conditioned 8D fit.
The solid lines indicate the selected DE solutions, and the dashed lines mark
the 16th, 50th, and 84th percentiles.
The off-diagonal panels show the strong correlations within the
effective-range sector and between the BDAD normalization and weak-current
coefficients.
Parameter units and marginal intervals are given in the corresponding tables
in Sec.~\ref{sec:results}.
}
\label{fig:azuma_corner}
\end{figure*}

\begin{figure*}[p]
\centering
\includegraphics[width=0.98\textwidth]{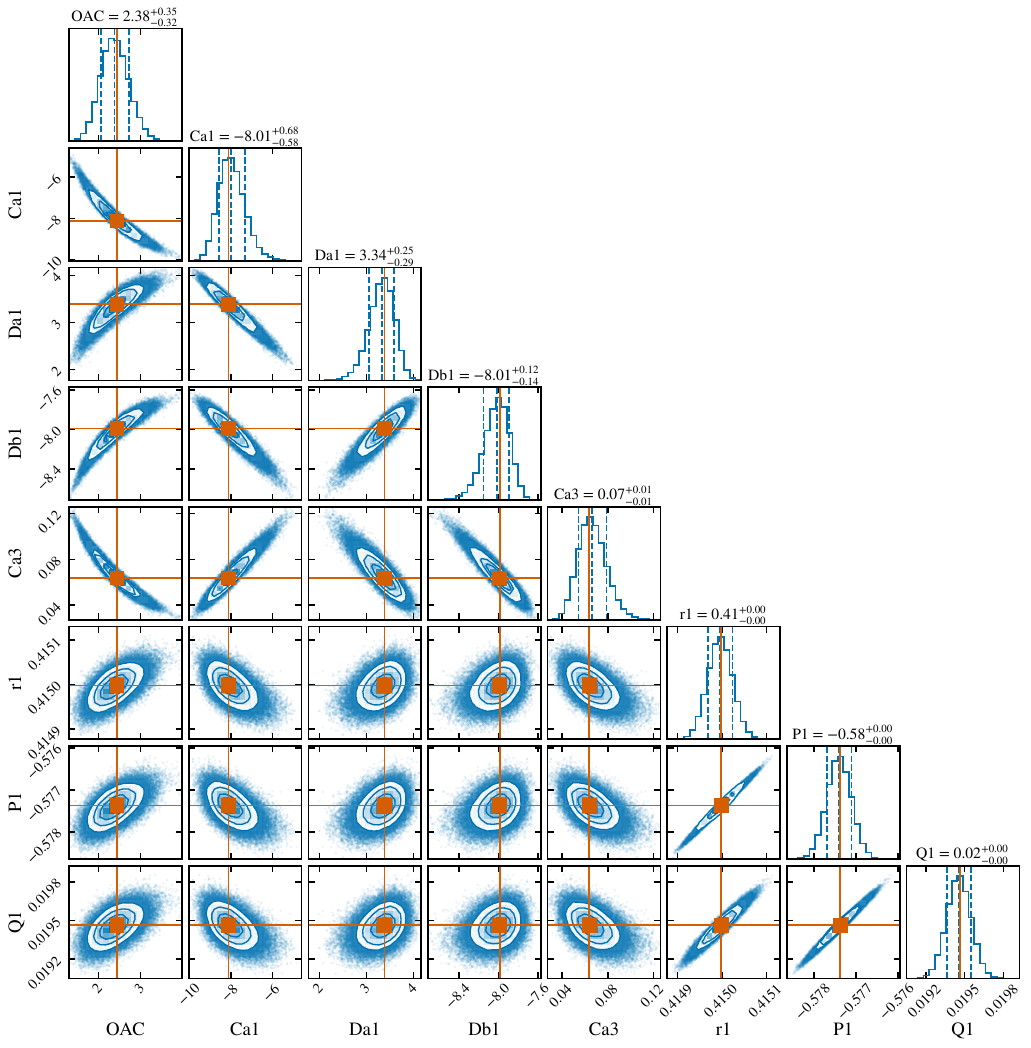}
\caption{
Pooled posterior distributions for the independent Tang-conditioned 8D fit.
See the caption of Fig.~\ref{fig:azuma_corner}.
As in Fig.~\ref{fig:azuma_corner}, the effective-range and weak-current
parameters form strongly correlated combinations.
The displayed posterior is conditional on the selected positive-
$C_a^{(3)}$ mode.
}
\label{fig:tang_corner}
\end{figure*}

Figures~\ref{fig:azuma_corner} and \ref{fig:tang_corner} show the full
marginal and joint posterior distributions for the independent Azuma- and
Tang-conditioned 8D fits.
The diagonal panels display the marginalized distributions, while the
off-diagonal panels show the corresponding two-dimensional posterior
densities.
The solid lines indicate the selected DE solutions, and the dashed lines mark
the 16th, 50th, and 84th percentiles.

\begin{table}[t]
\caption{
Selected Pearson correlation coefficients obtained from the pooled posterior
samples of the independent Azuma- and Tang-conditioned 8D fits.
}
\label{tab:independent_correlations}
\begin{ruledtabular}
\begin{tabular}{lcc}
Parameter pair
& Azuma
& Tang
\\
\hline
$\rho(r_1,P_1)$
& 0.994
& 0.992
\\
$\rho(P_1,Q_1)$
& 0.992
& 0.987
\\
$\rho(C_{\rm OAC},C_a^{(1)})$
& $-0.978$
& $-0.961$
\\
$\rho(C_a^{(1)},D_a^{(1)})$
& $-0.972$
& $-0.963$
\\
\end{tabular}
\end{ruledtabular}
\end{table}

Selected correlations from the independent fits are listed in
Table~\ref{tab:independent_correlations}.
These correlations show that the data constrain combinations of amplitudes
rather than isolated coefficients.
Marginal intervals and derived quantities, including the $1^-_1$ ANC, should
therefore be evaluated by propagating the full joint posterior rather than by
treating individual parameters independently.

For the Tang-conditioned 8D fit, 
the DE searches also identify a nearby second solution separated
from the selected minimum by 
$\Delta\mathcal J_{{\rm ind},T}=0.400$.
%
%
Its strong-sector parameters are nearly unchanged, whereas the sign of
$C_a^{(3)}$ is reversed.
The Tang posterior in Fig.~\ref{fig:tang_corner} and the parameter intervals
reported in 
Tables \ref{tab:weak_parameter_comparison} and \ref{tab:strong_diagnostics_8d},
especially the weak-current coefficients in Table~\ref{tab:weak_parameter_comparison},
are therefore conditional on the selected mode and
do not include this discrete ambiguity.

\begin{figure*}[p]
\centering
\includegraphics[width=0.90\textwidth]{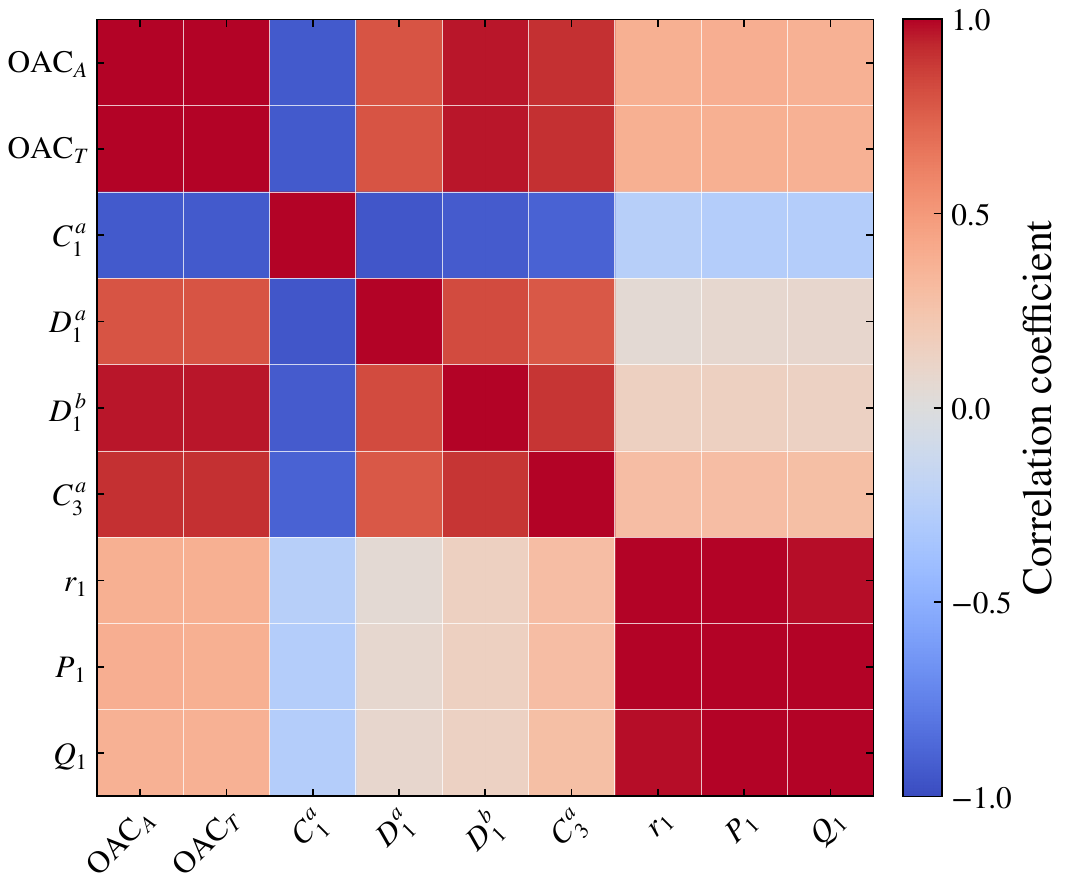}
\caption{
Posterior correlation matrix for the raw simultaneous 9D fit.
$C_{\rm OAC}^{A}$ and $C_{\rm OAC}^{T}$ are the independently fitted
Azuma and Tang normalization coefficients.
The remaining seven weak-current and effective-range parameters are common
to both BDAD spectra.
Selected correlation coefficients are listed in
Table~\ref{tab:simultaneous_correlations}.
}
\label{fig:joint_correlation}
\end{figure*}

\begin{figure*}[p]
\centering
\includegraphics[width=0.90\textwidth]{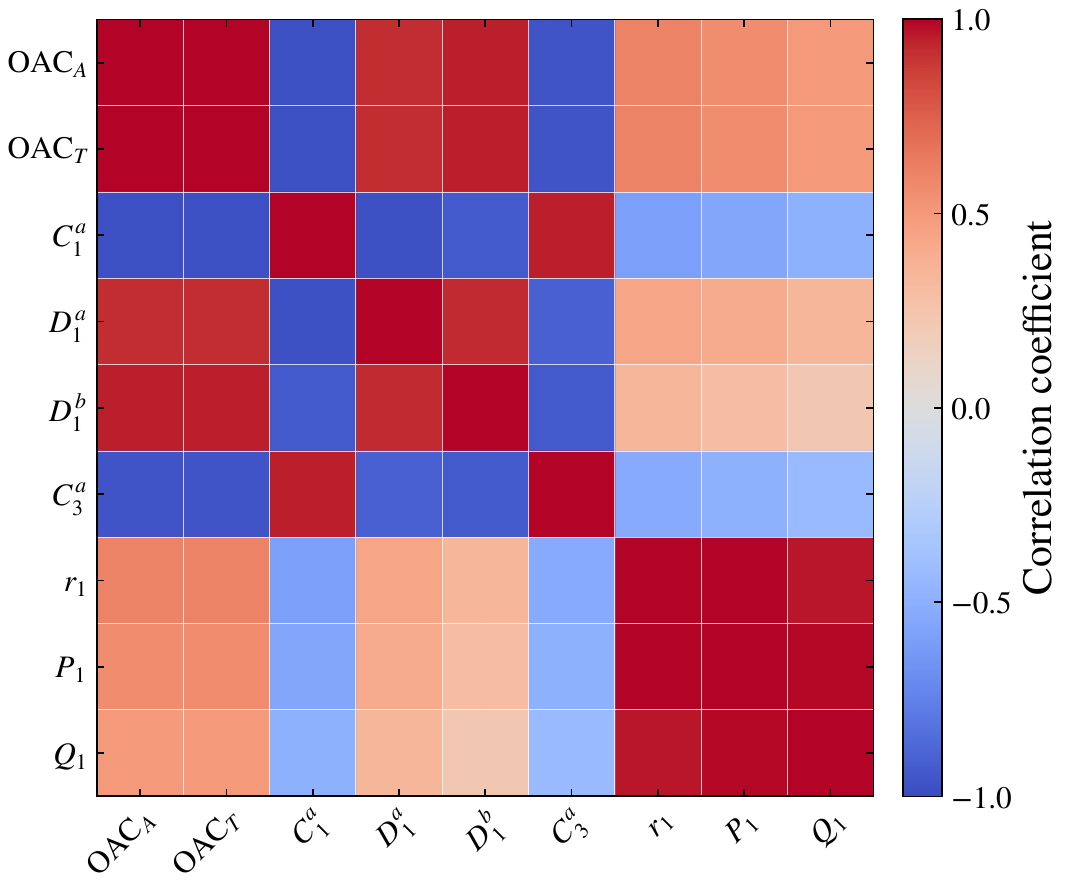}
\caption{
Posterior correlation matrix for the sector-balanced simultaneous 9D fit.
The parameter order and notation are the same as in
Fig.~\ref{fig:joint_correlation}.
Selected correlation coefficients are listed in
Table~\ref{tab:simultaneous_correlations}.
}
\label{fig:simultaneous_balanced_correlation}
\end{figure*}

Figures~\ref{fig:joint_correlation} and
\ref{fig:simultaneous_balanced_correlation} show Pearson correlation matrices
obtained from the pooled posterior samples of the raw and sector-balanced
simultaneous 9D fits, respectively.
Unlike the corner plots, these matrices provide a compact summary of linear
correlations rather than the full two-dimensional posterior shapes.
Thus, 
Figs.~\ref{fig:joint_correlation} and
\ref{fig:simultaneous_balanced_correlation}
are directly
comparable with each other, as are  Figs.~\ref{fig:azuma_corner} and \ref{fig:tang_corner}.
Comparisons between the two pairs are qualitative because the parameter
vectors and fitting objectives differ.

\begin{table}[t]
\caption{
Selected Pearson correlation coefficients obtained from the pooled posterior
samples of the raw and sector-balanced simultaneous 9D fits.
}
\label{tab:simultaneous_correlations}
\begin{ruledtabular}
\begin{tabular}{lcc}
Parameter pair
& Raw
& Sector-balanced
\\
\hline
$\rho(C_{\rm OAC}^{A},C_{\rm OAC}^{T})$
& 0.9985
& 0.9996
\\
$\rho(C_a^{(1)},D_a^{(1)})$
& $-0.9509$
& $-0.9719$
\\
$\rho(r_1,P_1)$
& 0.9957
& 0.9929
\\
$\rho(P_1,Q_1)$
& 0.9923
& 0.9901
\\
\end{tabular}
\end{ruledtabular}
\end{table}

Selected correlations from the two simultaneous analyses are compared in
Table~\ref{tab:simultaneous_correlations}.
The near-unit correlation between
$C_{\rm OAC}^{A}$ and $C_{\rm OAC}^{T}$ reflects a common
normalization--amplitude degeneracy induced by the shared weak-current
coefficients.
Both simultaneous objectives also retain strong correlations within the
$p$-wave weak-current and effective-range sectors.
The qualitative correlation pattern is therefore stable under the change of
objective, although the preferred weak-current parameter values are not.

The narrow simultaneous-fit intervals quantify the local posterior structure
within the selected common-amplitude mode.
They do not establish model adequacy, since both simultaneous fits retain
substantial BDAD line-shape residuals.
All intervals are conditional on the adopted parameter domains, fixed
branching-ratio inputs, fixed nonshared elastic parameters, and the pointwise
uncorrelated treatment of the experimental uncertainties.
The independent and sector-balanced simultaneous intervals are
generalized-posterior intervals conditional on their sector weights, whereas
the raw simultaneous intervals are based on the unweighted sum of the
pointwise sector chi-squares.

\begin{table}[t]
\caption{
Numerical convergence diagnostics for the simultaneous common-current 
fits.
}
\label{tab:simultaneous_numerics}
\begin{ruledtabular}
\begin{tabular}{lccc}
Analysis
& max $\widehat R$
& min $N_{\rm eff}$
& $N_{\rm post}/\max\tau$
\\
\hline
Raw
& 1.0033
& $1.99\times10^5$
& 292.0
\\
Sector-balanced
& 1.0082
& $3.42\times10^4$
& 160.8
\\
\end{tabular}
\end{ruledtabular}
\end{table}

The numerical diagnostics for the simultaneous analyses are summarized in
Table~\ref{tab:simultaneous_numerics}.
Independent DE searches reproduced the selected raw and sector-balanced
minima within $2.3\times10^{-7}$ and $3.1\times10^{-9}$ in their respective
objectives.
Both posterior calculations satisfy the convergence criteria adopted in
Sec.~\ref{subsec:protocol}.

\begin{table}[t]
\caption{
Uniform parameter domains used in the DE and MCMC analyses.
}
\label{tab:parameter_domains}
\begin{ruledtabular}
\begin{tabular}{lclc}
Parameter
& Range
& Parameter
& Range
\\
\hline
$C_{\rm OAC}^{A,T}\;(10^{\,6}\,\mathrm{MeV}^{-6})$
& $[0.05,50]$
&
$C_a^{(1)}\;(10^{-3}\,\mathrm{MeV}^{-2})$
& $[-50,20]$
\\
$D_a^{(1)}\;(\mathrm{MeV}^{-2})$
& $[-20,20]$
&
$D_b^{(1)}\;(10^{\,3}\,\mathrm{MeV})$
& $[-50,20]$
\\
$C_a^{(3)}\;(10^{-6}\,\mathrm{MeV}^{-4})$
& $[-1.0,0.5]$
&
$r_1\ ({\rm fm}^{-1})$
& $[0.39,0.44]$
\\
$P_1\ ({\rm fm})$
& $[-0.75,-0.40]$
&
$Q_1\ ({\rm fm}^{3})$
& $[0.005,0.040]$
\\
\end{tabular}
\end{ruledtabular}
\end{table}

The uniform parameter domains used in the DE and MCMC analyses are listed in
Table~\ref{tab:parameter_domains}.
The range assigned to $C_{\rm OAC}^{A,T}$ is also used for the single
normalization parameter in each independent fit.

\end{document}